\documentclass[aps,prl,preprint,groupedaddress,showkeys]{revtex4}
\usepackage{amsmath}
\usepackage{graphicx}
\begin{document}


\title{P-Stars}


\author{Paolo Cea$^{1,2,}$}
\email[]{Paolo.Cea@ba.infn.it}
\affiliation{$^1$Physics Department, Univ. of Bari, I-70126 Bari, Italy \\
$^2$INFN - Sezione di Bari, I-70126 Bari, Italy}



\begin{abstract}
P-Stars are a new class of compact stars made of up and down
quarks in $\beta$-equilibrium with electrons in an Abelian
chromomagnetic condensate. We show that P-Stars are able to
account for  compact stars with $R\lesssim 6 \, Km$, as well as
stars with radius comparable with canonical Neutron Stars. We find
that cooling curves of P-Stars compare rather well with
observational data. We suggest that P-Matter produced at the
primordial deconfinement transition is a viable candidate for
baryonic Cold Dark Matter. Finally, we show that P-Stars are able
to overcome the gravitational collapse even for mass much greater
than $10^6 M_{\bigodot}$.

\end{abstract}


\maketitle

One of the most intriguing aspect of modern astrophysics is the
possible existence of  Strange
Stars~\cite{alcock:1986,haensel:1986} which are made entirely of
deconfined  u, d, s quarks. The possible existence of Strange
Stars is a direct consequence of the so called Strange Matter
hypothesis, namely that  u, d, s quarks in equilibrium with
respect to weak interactions could be the true ground state of
hadronic matter~\cite{bodmer:1971,witten:1984}. \\
Recently, numerical lattice results  in
QCD~\cite{cea:2002a,cea:2002b} suggested that the gauge system
gets deconfined in strong enough Abelian chromomagnetic field
$A_{k}^a(\vec{x}) = \delta^{a,3} \, \delta_{k,2} \, x_1 \, H$.
This leads us to consider a new class of compact quark stars made
of almost massless deconfined up and down quarks immersed in a
chromomagnetic field in $\beta$-equilibrium. As we will discuss
below, it turns out that these compact stars, which will be
referred to as {\em P-Stars}, are more stable than Neutron Stars
whatever the value of  $\sqrt{gH}$, $g$ being the color gauge
coupling.
To investigate the structure of P-Stars we  needs the equation of
state appropriate for the description of deconfined quarks and
gluons in an Abelian chromomagnetic field. The quark chemical
potentials turn out to be smaller that the strength of the
chromomagnetic field  $\sqrt{gH}$.  Thus, we see that up and down
quarks are in the lowest Landau level with energy
$\varepsilon_{0,p_3} = |p_3|$, where $\vec{p}$ is the quark
momentum. It is now easy to find the thermodynamical potential (at
$T \, = \, 0$). We find for the energy density:
\begin{equation}
\label{equark}
 \varepsilon \; = \; \frac{1}{4 \pi^2} \, gH \,
 ( \mu_u^2 + \mu_d^2 )  \, + \, \frac{\mu_e^4 }{4 \pi^2} \,
 + \, \frac{11}{32 \pi^2} \, (gH)^2
 \, ,
\end{equation}
while the pressure P is given by:
\begin{equation}
\label{press}
 P \; = \; \frac{1}{4 \pi^2} \, gH \, ( \mu_u^2 +
\mu_d^2 ) \, + \, \frac{\mu_e^4 }{12 \pi^2}
 \, - \,  \frac{11}{32 \pi^2} \, (gH)^2 \, .
\end{equation}
where $\mu_f$ ( $f = d,u,e$) denote the chemical potential.
Moreover, we have the following constraints:
\begin{equation}
\label{constr1}
 \mu_e \, + \, \mu_u \; = \; \mu_d  \; \; \; \; \;
\; \;  \beta - equilibrium
\end{equation}
\begin{equation}
\label{constr2}
 \frac{2}{3 } n_u \, - \, \frac{1}{3 } n_d  \; = \;
n_e  \; \; \; \;  charge \; neutrality
\end{equation}
where:
\begin{equation}
\label{number}
  n_u \, = \, \frac{1}{2 \pi^2} \, gH \, \mu_u  \,
,  \;  n_d \, = \, \frac{1}{2 \pi^2} \, gH \, \mu_d  \, , \;  n_e
\, = \, \frac{\mu_e^3}{3 \pi^2} \, .
\end{equation}
We obtain the following Equation of State:
\begin{equation}
\label{eos}
 P \; = \; \varepsilon \;  -  \;  \frac{\mu_e^4 }{6
\pi^2}
 \; - \;  \frac{11}{16 \pi^2} \, (gH)^2 \, .
\end{equation}
It is interesting to observe that the sound speed:
\begin{equation}
\label{sound}
 v^2_S \, = \, \frac{d P}{d \varepsilon } \, = \,1 \, -  \;
\frac{1}{\frac{39}{2} \, + \, 18 \, \overline{\mu}^2_e \, + \,
\frac{15}{4 \, \overline{\mu}^2_e}} \; ,
\end{equation}
is quite close to the causal limit $v_S \, = \, 1$, for
$\overline{\mu}_e \, = \, \mu_e / \sqrt{gH} \, < \, 1 $, leading
to a very stiff equation of state. \\
To study some properties of a star we consider it to be a
spherical symmetric object, corresponding to a non rotating star.
The stability of the star is governed by the general-relativistic
equation of hydrostatic equilibrium for a spherical configuration
of quark matter, which is the Tolman-Oppenheimer-Volkov
equation~\cite{glendenning:2000}:
\begin{equation}
\label{TOV}
 \frac{d P}{d r} \, = \, - \; \frac{G M(r)
\varepsilon(r)}{r^2} \,[ 1 \, + \, \frac{4 \pi r^3 P(r)}{M(r)}] \;
\frac{1 \, + \,\frac{P(r)}{\varepsilon(r)} }{1 \, - \, \frac{2
GM(r)}{r}} \; ,
\end{equation}
\begin{equation}
\label{TOV1}
 \frac{d M}{d r} \, = \, 4 \; \pi \; r^2 \:
\varepsilon(r) \; .
\end{equation}
These equations can be solved numerically for a given central
density $\varepsilon_c$ and obtain $\varepsilon(r)$ and $P(r)$.
The radius of the star is determined by :
\begin{equation}
\label{radius}
 P(r=R) \; \; = \; \; 0 \; \; ,
\end{equation}
and the total mass by:
\begin{equation}
\label{mass}
  M \; \; =  \; \; M(r=R) \; \; \; .
\end{equation}
\begin{figure}
\includegraphics[width=0.8\textwidth,clip]{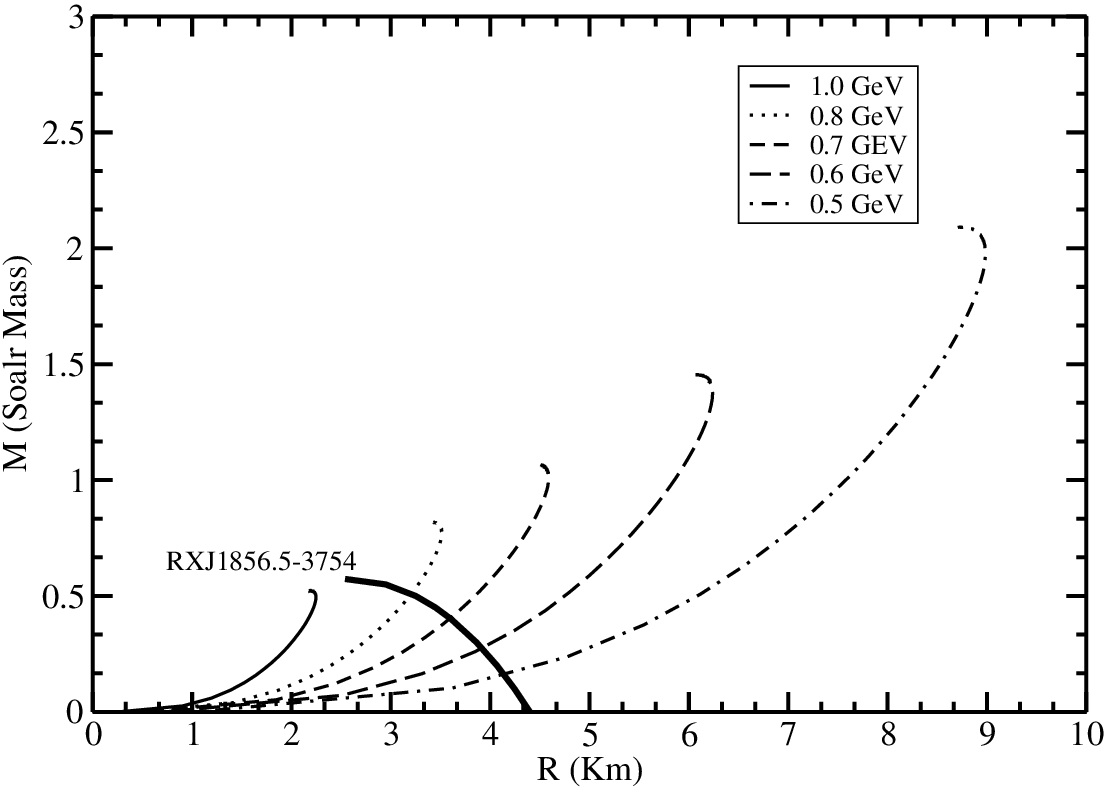}
\caption{\label{fig:01}
Gravitational mass M plotted versus stellar radius R for P-stars
for different values of $\sqrt{gH}$. Thick solid line corresponds
to M-R curve for {\it {RXJ1856.5-3754}} obtained  solving
Eq.~(\ref{radiuseff}) with $R^\infty = 4.4 \,Km$ and $R \geq
R_{\gamma} = 3 G M$ .}
\end{figure}
In Figure~1 we display the gravitational mass $M$ versus the star
radius $R$ for different values of $\sqrt{gH}$. At fixed value of
the chromomagnetic condensate the mass and radius of the star can
be thought of as function of the central energy density
$\varepsilon_c$. We see that, as in  Strange Stars, the mass first
increases with $\varepsilon_c$ until reaches a maximum value where
$\frac{d M(\varepsilon_c)}{d \varepsilon_c} \, = \, 0$. Increasing
further $\varepsilon_c$ leads in a region where $\frac{d
M(\varepsilon_c)}{d \varepsilon_c} \, < \, 0$ and the system
becomes unstable. Moreover, we see that there is no lower limit
for the radius R. Indeed, for small mass we find:
\begin{equation}
\label{radiuslow}
 R\, \sim \, M^{\frac{1}{3}} \; .
\end{equation}
Interestingly enough, Fig.~1 shows that there are stable P-Stars
with $M \, \lesssim \, M_{\bigodot}$ and $R \, \lesssim \, 6 \,
Km$. This region of the $M-R$ plane could be relevant for the
recently observed compact star {\it
{RXJ1856.5-3754}}~\cite{walter:1996,neuhauser:1997}. In fact, {\it
{RXJ1856.5-3754}} has been observed with Chandra and
XMM-Newton~\cite{burwitz:2002}, showing that the $X$-ray spectrum
is accurately fitted by a blackbody law. Assuming that the $X$-ray
thermal emission  is due to the surface of the star, the authors
of Ref.~\cite{burwitz:2002} found for the effective radius and
surface temperature:
\begin{equation}
\label{rad-temp}
 R^\infty \, \simeq \, 4.4 \, \frac{d}{120 \, pc} \, Km \; \; , \; \; T^\infty \, \simeq
 \, 63 \, eV \; ,
\end{equation}
where:
\begin{equation}
\label{radiuseff}
 R^\infty \, = \, \frac{R}{\sqrt{ 1 \, - \frac{2GM}{ R}}} \; , \;
 T^\infty \, = \, T \, \sqrt{ 1 \, - \frac{2GM}{ R}} \; .
\end{equation}
Assuming $ R^\infty = 4.4 \, Km$ (i.e. $ d = 120 \, pc$) we can
solve Eq.~(\ref{radiuseff}) for the true radius $R$. We also
impose that $ R \, \geq \, R_\gamma \, = \, 3 G M$, for it is well
known that the last circular photon orbit for the Schwarzschild
geometry is at $R_\gamma $. Remarkably, Fig.~1 shows that there
are stable P-Star configurations which agree with observational
data. However, it should be stressed that in the observed spectrum
there is an optical emission  in excess over the extrapolated
$X$-ray blackbody. Indeed, optical data are also fitted with a
blackbody model yielding an effective radius $R^\infty \, > \, 17
\, Km \, \frac{d}{120 \, pc}$\cite{burwitz:2002}. We feel,
however, that the seven data points in the optical range could be
interpreted differently. For instance,  the optical data could be
interpreted as  syncroton radiation emitted by electrons with
spectrum:
\begin{equation}
\label{syncroton}
N(E) = \kappa  E^{-\eta} , \, 2. \, 10^{-3} \, KeV \leq  E
 \leq \;7. \, 10^{-3} \, KeV  .
\end{equation}
We find a quite good fit with $\eta \simeq 0.71$. If this is the
case the radiation in the optical range should display a
polarization $ \Pi \, = \, \frac{\eta+1}{\eta+7/3} \, \simeq \, 0.56$. \\
Beside, there are stable P-Stars with mass and radius comparable
to those of Neutron Stars. In this case, we show that P-Stars are
more stable than Neutron Stars. To see this, we need the baryon
number:
\begin{equation}
\label{baryon}
A =  4 \pi \, \int_0^R \frac{n_b(r)}{\sqrt{ 1 \, - \frac{2GM(r)}{
r}}} \, r^2 \, dr \, ,
\end{equation}
where $n_b=\frac{1}{3}(n_u+n_d)$. Thus, the binding energy per
nucleon is :
\begin{equation}
\label{binding}
B \, = \, m_n \, - \,  \frac{M}{A} \, ,
\end{equation}
$m_n$ being the nucleon mass. We find for stars with limiting
mass:
\begin{equation}
\label{limiting}
\begin{split}
 B & \simeq 0.125 \; GeV \, , \, A \simeq 1.5 \, 10^{57} \, , \,
\frac{M}{M_{\bigodot}} \simeq 1.1  \\
B & \simeq 0.386 \; GeV \, , \, A \simeq 2.9 \, 10^{57} \, , \,
\frac{M}{M_{\bigodot}} \simeq 1.4 \, .
\end{split}
\end{equation}
These results should be compared with the binding energy per
nucleon in the case of limiting mass Neutron
Star~\cite{glendenning:2000}, $B_{NS} \simeq 100 \; MeV$, showing
that P-Stars are gravitationally more bounded than Neutron Stars. \\
Moreover, it turns out that P-stars are also more stable than
strange stars~\cite{alcock:1986,haensel:1986}. Indeed, it is known
that there is extra energy of about $20 \; MeV$ per baryon
available from the conversion of matter to strange
matter~\cite{alcock:1986}. So that, even the binding energy per
nucleon for strange stars, $B_{SS} \simeq 120 \; MeV$, is smaller
than the value in Eq.~(\ref{limiting}). This shows that
gravitational effects make two flavour quark matter in the
chromomagnetic condensate  globally more stable than nuclear
matter or three flavour quark matter. However, one could wonder if
locally, let's say in a small volume of linear size $a \; \simeq
1/m_\pi \; \simeq \; 1 \; fermi$, two flavor quark matter immersed
in the chromomagnetic condensate could convert into neutrons or
deconfined quarks of relevance for strange stars. In order to see
how far this conversion will proceed, we need to estimate the
appropriate transition rates. Let us first consider the decay into
nuclear matter. In this case we must rearrange three quarks, which
are in the lowest Landau levels, into configurations of quarks in
nucleons. In doing this, we get a first suppression factor
$\alpha$ due to color mismatch. In fact, the quarks into the
lowest Landau levels are colorless due to presence of the
chromomagnetic condensate. This suppression factor cannot be
easily evaluated. However, in general we have that $\alpha \;
\lesssim \; 1$. Another suppression factor arises from the
mismatch of the quark wavefunctions in the directions transverse
to the chromomagnetic field. This results in a factor $\left
(\frac{m_{\pi}}{\sqrt{gH}} \right )^2$ for each quarks. Finally,
to arrange three quarks into a nucleon we need to flip one quark
spin, which increases the energy by a factor $\sqrt{gH}$ at least.
So that we have for the transition rate:
\begin{equation}
\label{prob}
P \; \; \sim \; \; \alpha \; \left (\frac{m_{\pi}}{\sqrt{gH}}
\right )^6 \; \; e^{-\frac{\sqrt{gH}}{T}} \; \; ,
\end{equation}
where $T$ is the star core temperature. For a newly born compact
star the core temperature is of order of several $MeV$ and it is
rapidly decreasing. On the other hand, the typical values of
$\sqrt{gH}$ is of order of several hundred of $MeV$. As a
consequence we have:
\begin{equation}
\label{prob_num}
P \; \; \lesssim \; \; 1.5 \; \; 10^{-46} \; \;,
\end{equation}
which implies that the decays of two flavor quark matter  immersed
in the chromomagnetic condensate into nuclear matter is
practically never realized. In addition, the conversion into
three-flavor quarks is even more suppressed. Indeed, first we need
to flip a large number of quark spin to fill the Fermi sphere,
then the quarks at the top of the Fermi sphere are allowed to
decay into strange quarks. We conclude, thus, that P-stars, once
formed, are absolutely stable.

The binding energy is the energy released when the core of an
evolved massive star collapses. Actually, only about one percent
of the energy appears as kinetic energy in the supernova
explosion~\cite{bethe:1990}. Now, if we admit that  supernova
explosions give rise to P-Stars instead of Neutron Stars, then we
may completely solve the supernova explosion problems, for there
is an extra gain in kinetic energy  of about $ \, 1 \, - \, 10 \,
foe$ ($1 \, foe \, = \, 10^{51} \, erg$). Then we could eventually
 suppose that Pulsars are P-Stars and not Neutron Stars. Further
support to this point of view comes from cooling
properties of P-Stars. \\
Let us assume, for simplicity, stars of uniform density and
isothermal. The cooling equation is:
\begin{equation}
\label{cooling}
 C_V \; \frac{d T}{d t}
\; = \; - \; (L_{\nu} \; + \; L_{\gamma}) \; ,
\end{equation}
where $L_{\nu}$ is the neutrino luminosity, $L_{\gamma}$ is the
photon luminosity and $C_V$ is the specific heat. Assuming
blackbody photon emission from the surface at an effective surface
temperature $T_S$ we get:
\begin{equation}
\label{blackbody}
 L_{\gamma} \; = \; 4 \, \pi \, R^2 \, \sigma_{SB} \, T_S^4 \; ,
\end{equation}
where $\sigma_{SB}$ is the $Stefan-Boltzmann$ constant. Following
Ref.~\cite{shapiro:1983} we assume that the surface and interior
temperature are related by:
\begin{equation}
\label{surface}
 \frac{T_S}{T} \; = \; 10^{-2} \; a \; \; , \; \; 0.1 \; \lesssim a \; \lesssim 1.0
 \; .
\end{equation}
The dominant cooling processes by neutrino emission are the direct
$\beta$-decay quark reactions~\cite{iwamoto:1980,burrows:1980}:
\begin{equation}
\label{urca}
 d \; \rightarrow \; u \, + \, e \, + \, \overline{\nu}_e,\; \; , \; \;
u \, + \, e \, \; \rightarrow \; d \,  + \, \nu_e,\; \; .
\end{equation}
Let $|M|^2$ be the squared invariant amplitude averaged over
initial spin and summed over final spins. A standard calculation
gives:
\begin{equation}
\label{invariant}
 |M|^2 \; = \; 64 \, G_F^2 \cos^2\theta_c \, (p \cdot q') (p' \cdot q) \;
 ,
\end{equation}
where the quark momenta $p$ and $p'$ are directed along the
chromomagnetic field direction, $q$ and $q'$ are the electron and
neutrino momentum respectively . Now energy and momentum
conservation give:
\begin{equation}
\label{conservation}
\begin{split}
 p & \, = \, (E_p,p_3) \; , \; |p_3| \, \simeq \, \mu_d \; ; \; p' \, = \, (E_{p'},p'_3) \; , \;
  |p'_3| \, \simeq \, \mu_u  \\
 q & \, = \, (E_q,q_3,\vec{q}_{\bot}) \; , \;   |q_3| \, \simeq \, \mu_e \; , \;  |\vec{q}_{\bot}|
 \, \simeq \, T \\
 q' & \, = \, (E_{q'},\vec{q}') \; , \; |\vec{q}| \, \simeq \, T \; .
\end{split}
\end{equation}
So that we find:
\begin{equation}
\label{invariant2}
 |M|^2 \; \simeq \; 32 \, G_F^2 \cos^2\theta_c \, E_p \, E_{p'} \, E_q \, E_{q'} \;
 (\frac{T}{\mu_e})^2 \;  .
\end{equation}

Using standard methods for calculating phase space integrals in
degenerate Fermi systems~\cite{morel:1962}, we find the following
neutrino emissivity:
\begin{equation}
\label{emissivity}
 \epsilon_{\nu} \; \simeq \; \frac{1371}{30240} \; \sqrt{\frac{\pi}{2}} \; G_F^2 \; \cos^2\theta_c \;
 (\overline{\mu}_e)^2 \; \sqrt{gH} \;T^8
 \;  .
\end{equation}
Thus, the neutrino luminosity is:
\begin{equation}
\label{luminosity}
 L_{\nu} \; \simeq \; 3.18 \; 10^{36} \; \frac{erg}{s} \; T_9^8 \;
 \frac{M}{M_{\bigodot}} \; \frac{\varepsilon_0}{\varepsilon} \;
   \frac{\sqrt{gH}}{1 \, GeV} \;
 \;  ,
\end{equation}
where $T_9$ is the temperature in units of $10^9$ $\, {}^\circ K$,
and $\varepsilon_0 = 2.51 \; 10^{14} gr/cm^3$ is the nuclear
density. Finally, the specific heat is given by:
\begin{equation}
\label{specific}
 C_V \; \simeq \; 0.92 \; 10^{55} \;  T_9 \;
 \frac{M}{M_{\bigodot}} \; \frac{\varepsilon_0}{\varepsilon} \;
   (\frac{\sqrt{gH}}{1 \, GeV})^2 \; .
\end{equation}
\begin{figure}
\includegraphics[width=0.8\textwidth,clip]{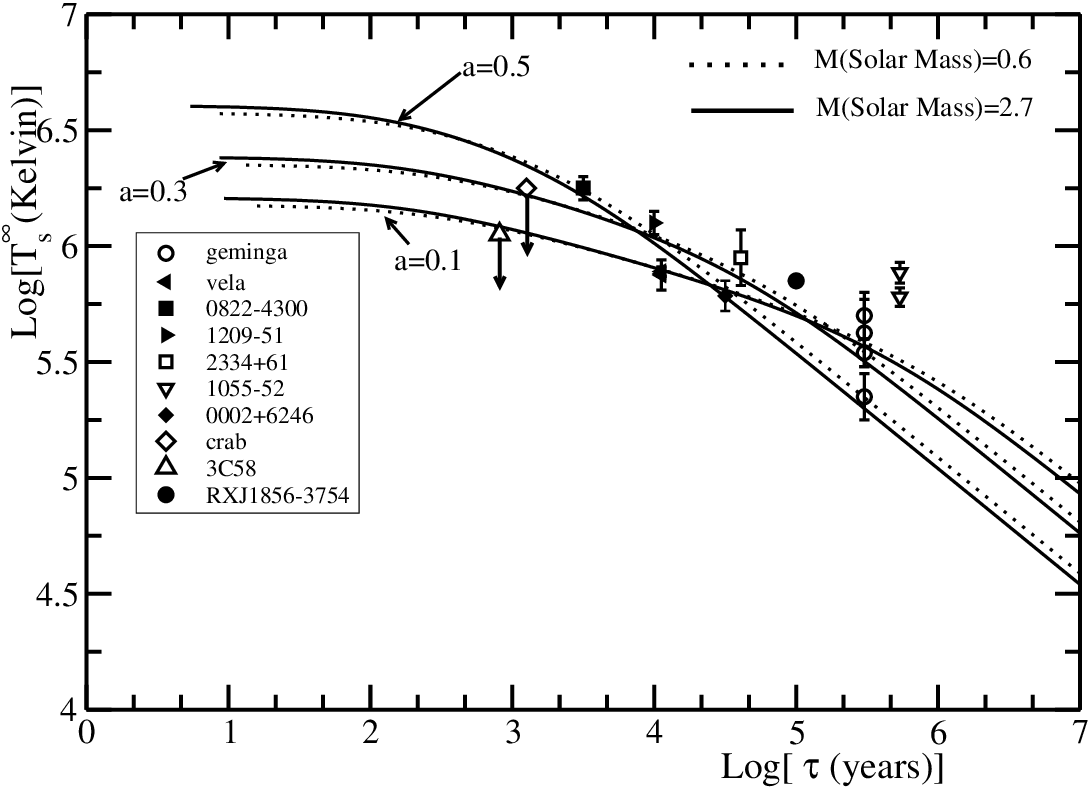}
\caption{\label{fig:02}
P-star cooling curves for $M = 0.6 \, M_{\odot}$ and $M = 2.7 \,
M_{\odot}$  and three different values of the parameter $a$ in
Eq.~(\ref{surface}). Experimental data for the effective surface
temperature have been taken from
Refs.~\cite{page:1997,burwitz:2002,slane:2002}. }
\end{figure}
We obtain the thermal history of a P-Star by integrating
Eq.~(\ref{cooling}) with $T^{(i)}_9 = 1.4$. \\
In Figure~2 we report our cooling curves for three different
values of the parameter $a$ in Eq.~(\ref{surface}). It is
worthwhile to stress that the cooling curves are almost
independent on the P-Star mass. Moreover, there is a weak
dependence on $a$ up to age $ \tau \, \sim \, 10^3 \, years $. We
compare our theoretical cooling curves with Pulsar data taken from
Ref.~\cite{page:1997}. We also report {\it {RXJ1856.5-3754}}
effective surface temperature assuming $\tau \, \simeq \, 10^5 \,
years$~\cite{burwitz:2002}, and the surface effective temperature
upper limit for supernova remnant {\it {3C58}}~\cite{slane:2002}.
We stress that  the upper limit has been obtained with $R^{\infty}
=  12 \; Km$. Assuming an effective radius smaller than the
Neutron Star canonical value should increase the above upper
limit.  In any case we see that the agreement between theoretical
cooling curves and observational data is quite satisfying.
Long time ago Witten~\cite{witten:1984} conjectured that quark
nuggets could be copiously produced during the cosmological
deconfinement phase transition, providing an explanation for the
Dark Matter in the Universe. Whether or not Strange
Matter~\cite{farhi:1984} is absolutely stable, the authors of
Ref.~\cite{alcock:1985} argued that a large lump of Strange Matter
evaporates only its outer layer, so that lumps large enough could
survive until today. In fact, a detailed
analysis~\cite{alcock:1985} showed that minimum baryon number of
lump that survives is roughly $ 10^{52}$, restricting severely the
viability of Strange Matter as a candidate of Dark Matter. On the
other hand, we now  show that this limitation is not shared by
P-Matter, i.e. lumps of up and down quarks in equilibrium with
respect to the weak interactions immersed in a chromomagnetic
condensate. \\
Recently, it has been suggested that strong magnetic fields of
order $10^{19} \, G$, corresponding to $\sqrt{gH} \sim 1 \, GeV$,
are naturally associated with the QCD scale~\cite{kabat:2002}.
Moreover, it is believed  that large magnetic fields might be
generated during cosmological phase
transitions~\cite{dolgov:2001}. So that it is conceivable that
P-Matter with chromomagnetic condensate $\sqrt{gH} \sim 1 \, GeV$
could have been copiously produced at the cosmological
deconfinement transition.
The physical properties of P-Matter are very similar to those of
Strange Matter. Let $A$ indicate the baryon number of a lump of
P-Matter, then we find:
\begin{equation}
\label{p-mass}
 M \, \simeq \, (2.5 \, GeV) \;  A \; , \; R \, \simeq \,
 \frac{A^{1/3}}{(\frac{4 \pi}{3} n_b)^{1/3}} \; ,
\end{equation}
where $n_b \simeq (248 \, MeV)^3$. Now, the baryon number of
P-Matter can change only through the emission or absorption of
neutrons. Moreover, we estimate the binding energy $I \, \gtrsim
\, \sqrt{gH}$. As for Strange Matter, the neutron cross section is
almost geometric:
\begin{equation}
\label{cross}
 \sigma \, \simeq \, 4 \, \pi \, R^2 \, f_n \, \simeq \, f_n \, \sigma_0 \, A^{2/3} \; ,
\end{equation}
where $\sigma_0 \simeq 0.79 \, 10^{-4} \, MeV^{-2}$ and
$f_n < \left (\frac{m_{\pi}}{\sqrt{gH}} \right )^2
 \simeq 0.04$.
 Following Ref.~\cite{alcock:1985} we find:
\begin{equation}
\label{rate1}
\frac{d A}{d t} \; = \; - \; r(A) \; , \; r(A) \; \simeq \;
\frac{m_n}{2 \pi^2} \, T^2 e^{-\frac{I}{T}} \, f_n \sigma_0
A^{2/3} \; .
\end{equation}
The evaporation rate of a lump of P-Matter when the Universe had a
temperature $T_U$ is given by:
\begin{equation}
\label{rateuniv}
\frac{d A}{d T_U} \; = \; \frac{d A}{d t} \; \frac{d t}{d T_U} \;
= \; \frac{d A}{d t} \; \sqrt{\frac{45}{172 \pi^3 G}} \;
\frac{-2}{T_U^3} \; .
\end{equation}
Integrating Eq.~(\ref{rateuniv}), assuming the conservative value
$f_n \simeq 0.1$, with typical $\sqrt{gH} \simeq 1.0 \, GeV$ and
initial temperature $T_U \simeq T_c \simeq 150 \, MeV$, we find
that the minimum size lump of P-Matter that survives until today
has a baryon number $A_{min} \sim 10^{42}$. This number should be
compared with $A_c$, the mean baryon number per horizon at $T_c$
assuming that the Universe is closed by baryons: $A_c \simeq
10^{48}$. We see that $A_{min}$ is much smaller than $A_c$, making
P-Matter a viable candidate for Cold
Dark Matter. \\
\begin{figure}
\includegraphics[width=0.8\textwidth,clip]{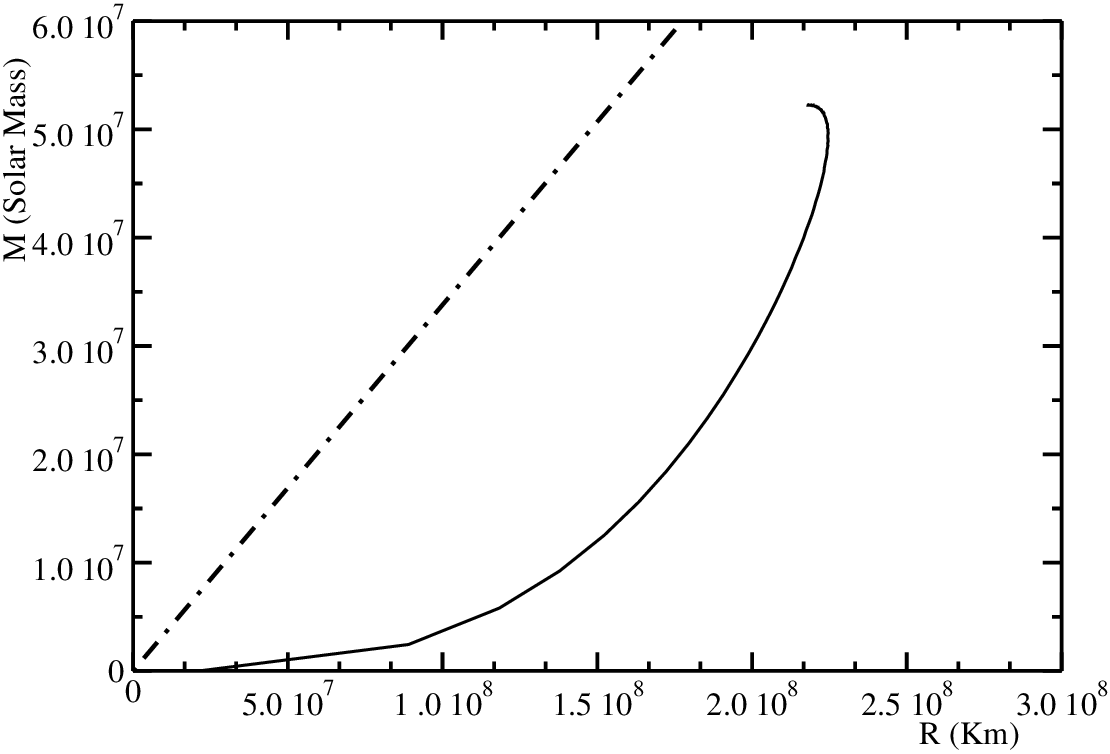}
\caption{\label{fig:03}
P-star gravitational mass M  versus stellar radius R for
$\sqrt{gH} = \, 10^{-4} \, GeV$. Dash-dotted line corresponds to
the Schwarzschild radius  $R_S \, = \, 2 G M$ .}
\end{figure}
The most interesting and intriguing aspect of our theory is that
P-Stars are able to overcome the gravitational collapse even for
mass much greater $10^6 M_{\bigodot}$. To see this, we note that
on dimensional ground we have:
\begin{equation}
\label{dimensional}
 M \; = \; \frac{1}{G^{3/2} gH} \; \;
 f(\overline{\varepsilon}_c) \;
\; \; \; \; \; R \; = \; \frac{1}{G^{1/2} gH} \; \;
g(\overline{\varepsilon}_c) \; ,
\end{equation}
where $\overline{\varepsilon}_c \, = \, \varepsilon_c/(gH)^2 $. As
a consequence we get:
\begin{equation}
\label{ratio}
\frac{2 \; G \; M}{R} \; = \; 2 \;
\frac{f(\overline{\varepsilon}_c)}{g(\overline{\varepsilon}_c)} \;
\; \equiv \; h(\overline{\varepsilon}_c).
\end{equation}
From Equation~(\ref{dimensional}) we see that by decreasing the
strength of the chromomagnetic condensate we increase the mass and
radius of the star. However, the ratio $\frac{2 \; G \; M}{R}$
depends only on  $\overline{\varepsilon}_c $. It turns out that
the function $h(x)$ defined in Eq.~(\ref{ratio}) is less than 1
for any allowed values of $\overline{\varepsilon}_c$. Indeed, in
Fig.~3 we shows stable configurations with $M \sim 10^7 \,
M_{\bigodot}$ and $R \gtrsim 1.3 \, R_S$, $R_S$ being the
Schwarzschild radius. Thus, we infer that our P-Stars do not admit
the existence of an upper limit to the mass of a completely
degenerate configuration. In other words, our peculiar equation of
state of degenerate up and down quarks in a chromomagnetic
condensate allows the existence of finite equilibrium states for
stars of
arbitrary mass. \\
Let us conclude by stressing the main results of this paper. We
found that P-Stars are able to describe very compact stars as well
as stars in the Neutron Star region. We showed that the cooling
curves compare reasonable well with observational data. Moreover,
we found that P-Matter could be a candidate for baryonic Cold Dark
Matter. However, the most striking result is that P-Stars are able
to halt the gravitational collapse. So that we may conclude that
P-Stars are challenging Neutron Stars and Black Holes.

\end{document}